\documentclass{emulateapj}

\def\lsim{~\raise0.3ex\hbox{$<$}\kern-0.75em{\lower0.65ex\hbox{$\sim$}}~}
\def\gsim{~\raise0.3ex\hbox{$>$}\kern-0.75em{\lower0.65ex\hbox{$\sim$}}~}

\def\gt{~\hbox{$>$}~}
\def\lbrack2{[\![}
\def\rbrack2{]\!]}

\def\nSrcInitial{{n_{\rm 0}^{\rm src}}}
\def\nSrcExpanded{{n_{\rm e}^{\rm src}}}

\def\i0g{{I_{0\rm g}}}
\def\e0g{{E_{0\rm g}}}

\def\kms{{\rm\,km\,s^{-1}}}
\def\kpc{{\rm\,kpc}}

\def\ev{{\rm\,eV}}
\def\msun{{\rm\,M_\odot}}

\def\pc{{\rm\,pc}}

\def\yr{{\rm\,yr}}

\def\Myrs{{\rm\,Myrs}}

\def\fesc{{f_{\rm esc}}}

\shorttitle{Escape of ionizing radiation from star forming regions}
\shortauthors{Razoumov \& Sommer-Larsen}

\begin{document}


\title{Escape of ionizing radiation from star forming
  regions in young galaxies}


\author{Alexei O. Razoumov\altaffilmark{1}}
\email{razoumov@ap.smu.ca}

\author{Jesper Sommer-Larsen\altaffilmark{2}}
\email{jslarsen@dark-cosmology.dk}


\altaffiltext{1}{Institute for Computational Astrophysics, Dept. of
  Astronomy \& Physics, Saint Mary's University, Halifax, NS, B3H 3C3,
  Canada}

\altaffiltext{2}{Dark Cosmology Centre, Niels Bohr Institute,
  University of Copenhagen, Juliane Maries Vej 30, DK-2100 Copenhagen,
  Denmark}


\begin{abstract}
  Using results from high-resolution galaxy formation simulations in a
  standard $\Lambda$CDM cosmology and a fully conservative
  multi-resolution radiative transfer code around point sources, we
  compute the energy-dependent escape fraction $\fesc$ of ionizing
  photons from a large number of star forming regions in two galaxies
  at five different redshifts from $z=3.8$ to 2.39. All escape
  fractions show a monotonic decline with time, from (at the
  Lyman-limit) $\sim6-10\%$ at $z=3.6$ to $\sim1-2\%$ at $z=2.39$, due
  to higher gas clumping at lower redshifts. It appears that increased
  feedback can lead to higher $\fesc$ at $z\gsim3.4$ via evacuation of
  gas from the vicinity of star forming regions and to lower $\fesc$
  at $z\lsim2.39$ through accumulation of swept-up shells in denser
  environments. Our results agree well with the observational findings
  of \citet{inoue..06} on redshift evolution of $\fesc$ in the
  redshift interval $z=2-3.6$.
\end{abstract}


\keywords{galaxies: formation --- intergalactic medium --- HII regions
  --- radiative transfer}

\section{Introduction}

In the redshift interval $2<z<6$ most ionizing photons in the Universe
are thought to originate in PopII stars in normal ($V_c\ga30-50\kms$)
galaxies \citep{nagamine......06}. For reionization calculations at
$z>6$ photons from lower mass halos should be taken into account
\citep{iliev...06}, whereas at $z\sim 2$ active galactic nuclei start
to play a dominant role \citep{madau..99}. The fact that the
metagalactic UV field peaks at $2<z<4$ \citep[and
ref. therein]{inoue..06} points to a peak in the galactic star
formation (SF) rate at such redshifts \citep[see
also][]{panter...06}. A major uncertainty in modeling the effect of
this SF on the thermal state of the intergalactic medium (IGM) is the
value of the escape fraction $\fesc$ of ionizing UV photons, which is
a function of redshift, and galaxy type and mass.

Best observational estimates of $\fesc$ come from the detailed
multiwaveband studies of the local Universe. \citet{leitherer...95}
find that less than $3\%$ of ionizing photons from low-redshift
starburst galaxies escape into the IGM. More recently, for several
local starburst galaxies, \citet{heckman.....01} estimate $\fesc\lsim
6\%$, noting that inclusion of dust will further reduce $\fesc$, and
\citet{bergvall..06} find $\fesc\sim 4-10\%$ for another local
starburst.

The situation is markedly different at $z\gsim3$, where the Lyman
continuum (LyC) detection points at much higher $\fesc$
\citep{steidel..01}. By comparing the direct observations of the LyC
from galaxies to the mean cosmic UVB intensity, \citet{inoue..06}
conclude that the average $\fesc$ increases from $1-2\%$ at $z=2$ to
$\sim10\%$ at $z\gsim3.6$.

Most theoretical estimates of $\fesc$ to date have been based on
models with a smooth distribution of gas \citep{ricotti.00} or
semi-analytical models with expanding shells and superbubbles in a
disk galaxy \citep{dove..00,fujita...03}. In this letter we present ab
initio calculations of the energy-dependent $\fesc$ of ionizing
radiation from SF regions in proto-galaxies found in numerical
simulations at five different redshifts from $z=2.39$ to 3.8. We
define $\fesc(\nu,r)$ simply as a fraction of photons of energy $h\nu$
which reach a shell of radius $r$ around a given source; by definition
$\fesc(\nu,0)=1$. We postprocess high-resolution simulation datasets
with radiative transfer (RT) on nested grids, with the maximum grid
resolution of $15\pc$. Each galaxy contains from several hundred to
several thousand distinct stellar sources representing individual,
young SF regions. Our interstellar gas distribution is extremely
clumpy due to an array of physical processes including feedback from
SF. We also estimate the effect of the dynamical expansion of HII
regions on the escape of ionizing photons.

\section{The code and models}

\begin{figure}
  \epsscale{.60}\plotone{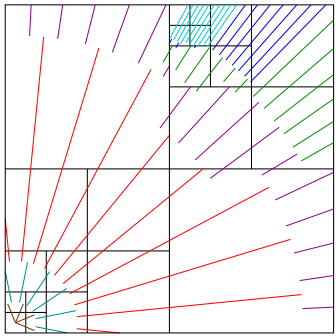}
  \caption{Radial ray geometry in the FTTE point source
    algorithm. Color represents the angular resolution level. To avoid
    a non-uniform coverage of a cell such as that seen at the top of
    this box, we normally avoid cell interfaces with a jump in
    refinement by more than one level.}
  \label{radialGeometry}
\end{figure}

To compute the escape of ionizing radiation from SF regions, we use an
extension of the Fully Threaded Transport Engine
\citep[FTTE,][]{razoumov.05} to point sources. This new module was
previously briefly introduced in the comparison study by
\citet{iliev...............06}. It extends the adaptive ray-splitting
scheme of \citet{abel.02} to a model with variable grid
resolution. Sources of radiation can be hosted by cells of any level
of refinement, although usually in cosmological applications sources
are sitting on the deepest level of refinement
(Fig.~\ref{radialGeometry}). Around each point source we build a
system of $12\times4^{n-1}$ radial rays which split either when we
move further away from the source, or when we enter a refined cell,
and $n=1,2,...$ is the angular resolution level. Once a radial ray is
refined it stays refined, even if we leave the high spatial resolution
patch, until even further angular refinement is necessary. All ray
segments are stored as elements of their host cells, and for actual
transport we just follow these interconnected data structures
accumulating photo-reaction number and energy rates in each cell. With
multiple sources, we add the rates from all sources before updating
the time-dependent rate network using the chemistry solver from
\citet{abel...97}, and the global timestep is chosen such that each
species abundance does not change by more than $30\%$.



With a small ($N_{\rm src}<10^3$) number of star particles we can
compute a full ray geometry separately for each source. For a large
number of sources ($10^3-10^6$) the algorithm allows construction of
trees of sources in which individual tree nodes are treated as a
single source far from their origin.

Our algorithm shares many common ideas with the recent RT module of
the cosmological structure formation code Enzo implemented by
\citet{abel..06} to study feedback from Pop III stars and, in fact, is
a basis of an independently developed radiation hydrodynamics module
of Enzo which we will introduce elsewhere.

\subsection{Galaxy formation models}

We use results of high-resolution galaxy formation simulations in a
standard $\Lambda$CDM cosmology done with a significantly improved
version of the TreeSPH code described by \citet{sommer-larsen..03} --
some detail is given in \citet{sommer-larsen06}.  The simulations
invoke the formation of discrete star ``particles'', which represent a
population of stars born at the same time in accordance with a given
initial mass function (IMF). For the purpose of this work we focus on
two (proto-) galaxies, K15 and K33, which at $z$=0 become typical disk
galaxies of $V_c$= 245 and 180 $\kms$, respectively. The RT
calculation are performed at five different redshifts, $z=3.8$, 3.6,
3.4, 2.95, and 2.39, for ``normal'' resolution simulations, and at the
first three for ``high'' resolution ones (8 times higher mass
resolution, twice better force resolution). For galaxy K15 we also
vary the strength of supernova feedback (Table 1). Total particle
numbers range from $\sim1.6\times10^5$ to $\sim2.2\times10^6$; star
(and SPH) particle masses are $1.1\times10^6$ and $1.4\times10^5$
$\msun$, for normal and high resolution simulations, respectively.


\begin{table}
  \begin{center}
    \caption{Number of distinct star-particles of age$\le$34
      Myr.\label{table1}}
    \begin{tabular}{lccccccc}
      \tableline
      Simulation & resolution & feedback & z=3.8 & 3.6 & 3.4 & 2.95 & 2.39\\
      \tableline
      \multicolumn{8}{c}{galaxy 1}\\
      \tableline
      K15-1-8 & normal & $>$normal && 347 && 1000 & 560 \\
      K15-06-8 & normal & normal & 431 & 560 & 595 & 1161 & 596 \\
      K15-06-64 & high & normal & 5061 & 5248 & 6005 &&\\
      \tableline
      \multicolumn{8}{c}{galaxy 2}\\
      \tableline
      K33-06-8 & normal & normal & 306 & 340 & 390 & 337 & 144 \\
      K33-06-64 & high & normal & 2566 & 3122 & 3389 &&\\
      \tableline
    \end{tabular}
  \end{center}
\end{table}


Radiative transfer is computed on top of a nested data structure
containing 3D distributions of physical variables. To create such a
structure from the SPH simulation datasets, for each galaxy we cut out
a $(250\kpc)^3$ box centered on the galaxy and projected this box onto
a $128^3$ uniform grid. We then subdivided every base grid cell which
contains more than $N_{\rm max}=10$ SPH particles, and continued this
process of subdivision recursively so that no cell contains more than
$N_{\rm max}$ gas particles.

\subsection{Stellar population synthesis}

\begin{figure}
  \epsscale{1.05}\plotone{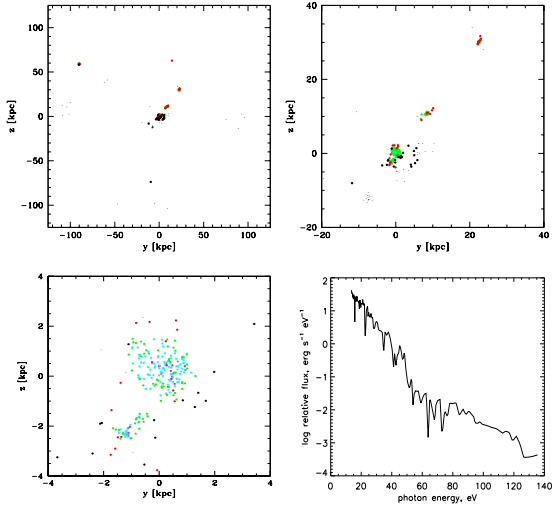}
  \caption{The distribution of all stellar particles (SF regions)
    younger than $t_{\rm up}=34\Myrs$ in the volumes 250\kpc (top
    left), 60\kpc (top right) and 8 kpc (lower left) on a side
    centered on the galaxy K15-1-8 at $z=3.6$. The color of each
    particle represents the level of the host cell: purple circles
    (level 5), light blue circles (level 4), green circles (level 3),
    red circles (level 2), black circles (level 1), and black dots
    (base grid level 0 cells). The lower right panel shows the input
    spectrum of a star particle.}
  \label{three}
\end{figure}

In the (normal resolution) galaxy K15-1-8 at $z=3.6$ there are 347
stellar sources younger than $t_{\rm up}=34\Myrs$
(Fig.~\ref{three}). The birth times of particles are distributed
almost uniformly over the span of $34\Myrs$, corresponding to a nearly
constant star formation rate (SFR) of $11.6\msun/\yr$. To compute the
stellar UV luminosity function for this and other galaxies, we use a
population synthesis package Starburst 1999
\citep{leitherer........99} with continuous SF at a constant rate
$11.6\msun/\yr$ distributed among 347 stars. For other galaxies and at
other redshifts the SFR per particle is adjusted to account for the
actual number of stars produced in the past $34\Myrs$, and the
specific (per unit spectrum) luminosity is distributed uniformly among
all stars in the volume.

We assume a triple-interval Kroupa IMF with indices (0.5,1.2,1.7)
above the mass boundaries $(0.1,0.5,1)\msun$, respectively, a
supernova cut-off at $8\msun$, a black hole cut-off $120\msun$, and
solar metallicity. The resulting spectrum is shown in the lower right
panel of Fig.~\ref{three}.


\subsection{Heating and HII region expansion}

The TreeSPH code assumes the UVB of Haardt \& Madau (1996) with
self-shielding in regions in which the mean free path of Lyman-limit
photons is less than 1\kpc. In addition to radiative heating and
cooling and the hydrodynamical $PdV$ term, some heating originates
from shock dissipation which we include in the following way.
After we project the 3D hydro fields onto the grids, in each cell we
apply the uniform UVB with the self-shielding correction and compute
the amount of additional heating needed to keep the temperature of
that cell constant (thermal equilibrium without sources). We then
switch on the stellar sources retaining the extra heating term, and
evolve simultaneously the equations of point source transfer and
non-equilibrium chemistry to $10\Myrs$. With continuous SF and
constant UV luminosities,
all $\fesc$ reach a plateau after the first few \Myrs. In this paper
all results are presented at $t=10\Myrs$ at which point we assume
convergence.


Direct momentum transfer from UV photons to the
interstellar/intergalactic gas is negligible
\citep{spitzer78,whalen.06}. However, heating and ionization by
stellar photons creates pressure gradients that drive expansion of the
HII regions into the surrounding gas. The physics of expansion has
been studied in detail in 1D \citep{kitayama...04,whalen.06}, 2D
\citep{shapiro..04}, and recently 3D \citep{abel..06} radiation
hydrodynamics simulations. Without coupling RT to the hydro code, we
cannot model this expansion self-consistently. However, we can mimic
its effect by lowering the density in the giant HII region around the
starburst region which will ease the escape of UV photons. We use the
results from section 3.1 of \citet{larsen..01} which lists the
"initial Str\"omgren radius" $R_0$ (1 Myr after the burst) and the
"final Str\"omgren sphere" $R_e$ (4 Myr after the burst; at this time
the UV luminosity declines rapidly) as a function of the local density
and the ionizing source luminosity.

For each source we use the initial gas density $\nSrcInitial$ of the
its host cell to find $R_0$ and $R_e$ and then modify the density of
each cell inside $R_e$ by $\nSrcExpanded/\nSrcInitial$, provided that
the initial density of that cell $n_{\rm 0}\le\nSrcInitial$. For
$n_{\rm 0}\gt\nSrcInitial$ the expanding shell hits a denser region,
and without a full hydrodynamical calculation one cannot predict
whether this will lead to expansion or compression of the denser
region, which explains our condition. Therefore, for a given source
our density correction is anisotropic which seems entirely realistic
for an HII front expansion in an inhomogeneous medium.


For such large galaxies, we do not expect our results to change
significantly if a fully self-consistent hydro/RT approach is used, as
the energetic and momentum effects of ionization fronts are negligible
compared to the effects of SNII explosions, shocks etc. which are
already included into our models, resulting in a very clumpy
interstellar medium (ISM). However, the radiative feedback would also
smooth some of the small-scale structure by raising the Jeans mass
which is especially important in lower mass galaxies, an effect which
we plan to include in our future simulations.

\section{Results and discussion}

\begin{figure}
  \epsscale{1.}\plotone{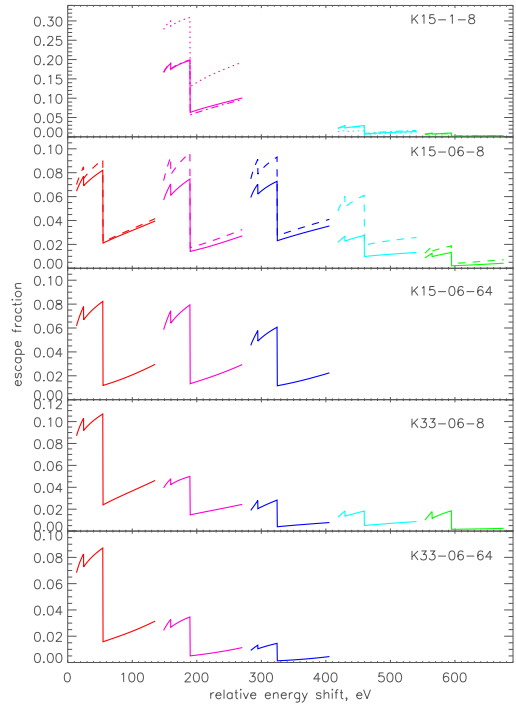}
  \caption{Spectral dependence of $\fesc$ at $r=100\kpc$ for all five
    input galaxy models at (where transfer has been computed) $z=3.8$
    (red), $z=3.6$ (magenta), $z=3.4$ (blue), $z=2.95$ (cyan), and
    $z=2.39$ (green). For each model the redshift evolution is from
    left to right, and each curve goes from $13.6\ev$ to $135\ev$. All
    solid lines are models without the HII region expansion, with the
    SFR distributed uniformly among all stars younger than
    $34\Myrs$. For K15-06-8 the dashed lines show models with the HII
    region expansion. For K15-1-8 we also computed the models with the
    {\it same} SFR distributed among stars younger $3.4\Myrs$ (dotted)
    and $10\Myrs$ (dash-dotted lines).}
  \label{escapeFraction}
\end{figure}

Fig.~\ref{escapeFraction} shows spectral dependence of $\fesc$ at
$r=100\kpc$ in each model, averaged over all sources. All $\fesc$ have
been corrected for the boundary of the volume. For individual sources
the escape of photoionizing radiation is similar to a phase
transition: it is very sensitive to conditions at the source and
therefore can vary by a sizable factor for the same galaxy over the
course of its evolution. We find that while there are variations among
the models, the escape fractions tend to decrease with lower redshift,
reflecting the fact that with time more gas cools into the higher
density clouds, as well the declining SFR at $z<2.95$. With normal
feedback in both galaxies $\fesc$ at the Lyman edge drops from
$8-10\%$ at $z=3.8$ to $\sim1-2\%$ at $z=2.39$.


Not surprisingly, in all our models $\fesc$ of photons capable of
doubly ionizing helium is very low, due to the softness of stellar
radiation. The escape fraction of photons capable of single He
ionization is comparable at all redshifts to that of H-ionizing
photons.


Increasing SN feedback in K15 evacuates more gas from the vicinity of
the SF regions leading to a threefold rise in $\fesc$ at
$z=3.6$. Lower local gas densities translate into a $40\%$ reduction
in the SFR (Table 1). At $z=2.95$, a denser environment near the
sources produces a much smaller difference in the SFR and $\fesc$, when
we increase the strength of feedback, and, at $z=2.39$, $\fesc$ actually
drops by $25\%$ as a larger fraction of gas swept-up by the shocks
stays in the vicinity of the SF regions.


Next we compare our normal and high resolution results, at $z=3.8-3.4$,
and find a good agreement between the two. We see it as an additional
test of our algorithm, since other properties of the two sets of
simulations (SFR, etc.) also match. 

Since our choice of distributing SF over the $t_{\rm up}=34\Myrs$
youngest stars is somewhat arbitrary, in galaxy K15-1-8 we also
experimented with the same amount of SF distributed over all stars
younger than $t_{\rm up}=10\Myrs$ and $3.4\Myrs$. Ideally, these young
stars would produce most of the ionizing UV photons. Since we do not
calculate transfer and hydro simultaneously, we feel that applying
continuous SF with the constant rate is a better computational
approach than assuming instantaneous SF which will have no lasting
effect on the thermal state of the IGM after few Myrs. With continuous
SF we expect our results to converge as we increase the number of star
particles. The results at $t_{\rm up}=10\Myrs$ and $34\Myrs$ are in
fact very close to each other at all three redshifts
(Fig.~\ref{escapeFraction}), whereas the $3.4\Myrs$ results are
somewhat off, likely related to the small number of sources used
to represent the SF in this latter case.

The dashed lines in K15-06-8 in Fig.~\ref{escapeFraction} show the
energy-dependent $\fesc$ at $r=100\kpc$ corrected for the HII region
expansion. At most redshifts expansion raises $\fesc$ by $10-30\%$,
although at $z=2.95$ $\fesc$ jumps by more than a factor of
two. Further analysis shows that most of this change can be attributed
to removal of the gas from the highest refinement level cells which
are hosting the sources, not the surrounding cells. 

\begin{figure}
  \epsscale{1.1}\plotone{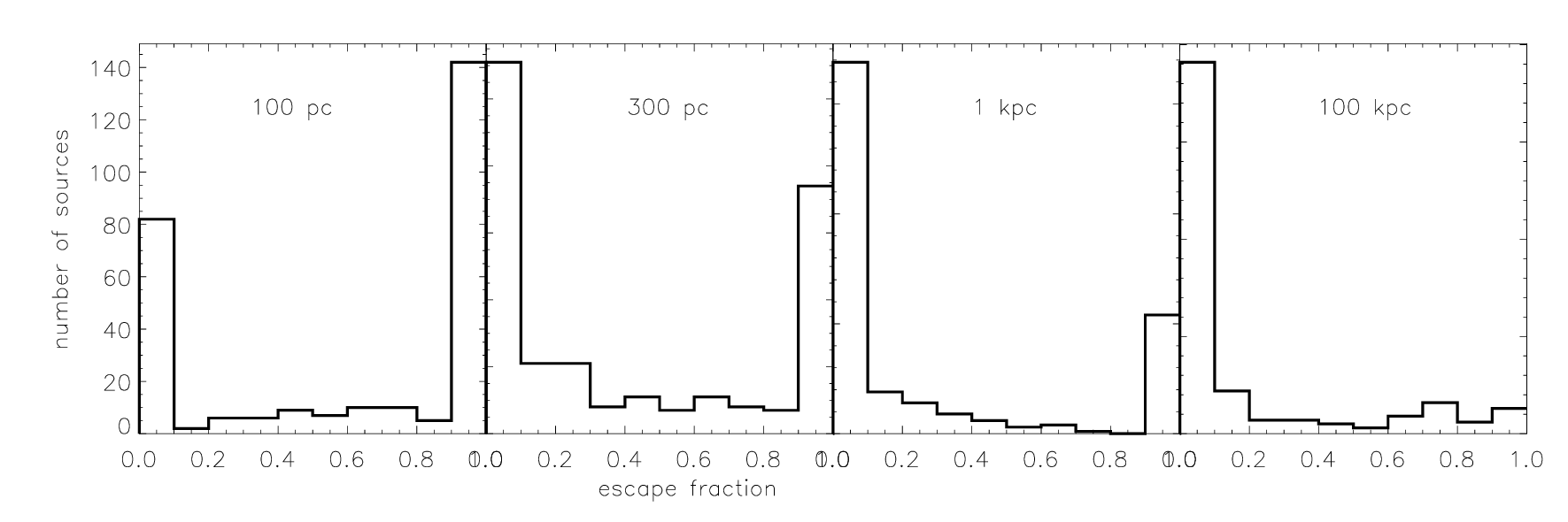}
  \caption{Lyman-limit escape fraction distribution for all sources in
    galaxy K15-1-8 at $z=3.6$ (without HII region expansion), at four
    different radii, from 100 pc to 100 kpc.}
  \label{radial}
\end{figure}

In Fig.~\ref{radial} we show how $\fesc$ changes with distance from
sources in galaxy K15-1-8 at $z=3.6$. A large fraction of photons
reaches the radius $r=100\pc$. As we do not have any reliable
information on gas clumping on sub-resolution scales at these
redshifts, the effective radius of absorption is not firmly
established at this point. In our current models most absorption
occurs within few hundreds pc of each star particle, well inside the
virial radius of $\sim$45 kpc. At $z=3.6$, the proto-galaxy consists of 
several star forming regions as well as a significant amount of the 
intergalactic HI, all of which account for some fractional absorption 
beyond $1\kpc$.


In this study we neglected the effect of dust
extinction. \citet{benson...06} point out that in the case of a smooth
ISM attenuation by dust can reduce $\fesc$ by a factor of few. For a
clumpy ISM at high redshifts the magnitude of dust attenuation should
be much lower, particularly that most ionizing photons are produced in
star bursts, as opposed to a quiescent disk.

In conclusion, our models with exact ionizing continuum RT around a
large number of SF regions in young protogalaxies show a monotonic
decline in $\fesc$ of ionizing photons from higher to lower
redshifts. With the normal feedback strength $\fesc$ at the Lyman edge
drops from $\sim6-10\%$ at $z=3.6$ to $\sim1-2\%$ at $z=2.39$. At
higher redshifts SF on average occurs at slightly lower densities
resulting in easier escape of UV photons into the IGM. This result
agrees well with the observational findings of \citet{inoue..06} on
redshift evolution of $\fesc$. Note that although the two galaxies in
our study have a mass ratio of $\sim2.5$, their $\fesc$ are very
similar. One could suggest that perhaps, once star formation begins,
it is the physical conditions in the clumpy gas clouds that determine
the escape of UV photons, rather than the overall properties of their
host galaxies. However, one would need a much larger statistical
sample to test this speculation. In our next paper we will further
examine the role of the fine structure of the clumpy ISM in the escape
of ionizing photons, as well as the importance of hydrodynamical
effects.


\acknowledgments

AR would like to thank Mika Juvela and Eric Lentz for valuable
discussions on numerical methods, as well as the hospitality of the
Dark Cosmology Centre at the University of Copenhagen. The TreeSPH
simulations were performed on the SGI Itanium II facility provided by
DCSC. The Dark Cosmology Centre is funded by the DNRF.

\bibliographystyle{apj}

\clearpage


\clearpage


\end{document}